\documentclass[%
    preprint,
    superscriptaddress,
    amsmath,
    amssymb,
    aps,
    pre,
    table, 
    xcdraw
]{revtex4-1}

\usepackage{amssymb}
\usepackage{setspace}
\usepackage[utf8]{inputenc}
\usepackage{array}
\usepackage{natbib}
\usepackage{amsmath}
\usepackage{float}
\usepackage{multirow}
\usepackage{hyperref}
\usepackage{rotating}
\usepackage{adjustbox}
\usepackage{xcolor}
\usepackage[utf8]{inputenc}
\usepackage[T1]{fontenc}
\usepackage{textcomp}
\usepackage{listings}
\usepackage{xcolor}
\usepackage[most]{tcolorbox}
\usepackage{fancyvrb}
\usepackage[most]{tcolorbox}
\usepackage{enumitem}

\tcbset{
  promptbox/.style={
    enhanced jigsaw,
    breakable,
    colback=gray!4,
    colframe=black!20,
    boxrule=0.4pt,
    arc=1.5mm,
    left=2mm,
    right=2mm,
    top=1mm,
    bottom=1mm,
    before skip=10pt,
    after skip=10pt
  }
}

\begin{document}

\title{From Node2Vec to GPT-based GraphRAG: scientific impact prediction across graph and language models}

\author{Adilson Vital Jr.}
\affiliation{Institute of Mathematics and Computer Sciences,
Universidade de S\~ao Paulo, PO Box 369,
13560-970, S\~ao Carlos, SP, Brazil 
}

\author{Filipi N. Silva}
\affiliation{Northwestern University, Evanston, IL, USA}
\affiliation{Indiana University, Bloomington, IN, USA}

\author{Diego R. Amancio}
\affiliation{Institute of Mathematics and Computer Sciences,
Universidade de S\~ao Paulo, PO Box 369,
13560-970, S\~ao Carlos, SP, Brazil 
}

\date{\today}

\begin{abstract}
Identifying which newly published scientific papers are likely to become highly cited is important for prioritizing research attention, supporting editorial decisions, and guiding the allocation of scientific resources, particularly under cold-start conditions where little direct evidence is available at publication time. In this work, we formulate impact prediction as a cohort-normalized top-$P\%$ classification task and compare graph-based and LLM-based approaches under a unified framework. 
We construct citation and textual-similarity graphs under temporal constraints and generate Node2Vec representations, either alone or combined with OpenAI text embeddings. The best supervised configuration combines directed citation graphs with textual embeddings, reaching approximately 0.84--0.85 AUC. We also evaluate a GPT-based GraphRAG setup, using GPT 5.5 and 5.4 Nano, in which graph neighborhoods are used as contextual evidence for prediction. Although the LLM-based approach achieves high performance, retrieved context does not consistently improve results; target-only prompts often perform as well as or better than GraphRAG prompts achieving the 0.87 mark. These findings indicate that structural and textual signals are complementary for supervised prediction, while retrieval augmentation must be carefully evaluated against simpler LLM baselines.
\end{abstract}

\maketitle

\section{Introduction}
\label{chapter:introduction}

Predicting the future impact of recently published papers is a central problem in the science of science. Beyond its theoretical relevance for understanding how scientific influence emerges and diffuses, it also has important practical applications in research assessment, editorial triage, journal screening, and the early identification of promising topics and contributions \cite{kousha2024factors,STEGEHUIS2015642,ABRAMO201932}. Although scientific impact can be operationalized in multiple ways, citation counts remain one of the most widely adopted indicators of scientific impact and relevance, albeit with well-known conceptual and methodological limitations \cite{aksnes2019citations,waltman2016review}. In this context, impact is often more meaningfully assessed in relative rather than purely absolute terms, comparing a paper's citation performance with that of an appropriate peer cohort under similar temporal conditions; percentile-based indicators provide one established way of doing so \cite{waltman2013percentile,waltman2016review}.

This task is inherently difficult because scientific impact is driven by a heterogeneous interplay of textual, bibliometric, relational, and social factors that vary across disciplines and timeframes~\cite{kousha2024factors}. 
Newly published papers are particularly challenging to evaluate as they face a strict cold-start problem, characterized by a near-total absence of direct evidence regarding their future citation trajectory. Consequently, predictive models must rely on information available at the point of publication, such as textual content, authorship metadata, and citation-based relational signals \cite{STEGEHUIS2015642,ABRAMO201932}. However, these data sources are heterogeneous and partially observable, often proving difficult to integrate into structured representations suitable for modeling. Furthermore, the exponential growth in scientific output and citation volume induces a temporal inflation in raw citation counts; this renders direct cross-year comparisons misleading unless temporal variables are strictly controlled \cite{petersen2019inflation,waltman2016review}. Effective impact prediction, therefore, necessitates representations that simultaneously capture a paper’s semantic essence and its placement within the evolving topology of citation networks. Such integration is achievable by modeling papers through complex networks, a methodology that has been successfully applied in other domains to study the evolution of complex entities~\cite{brito2020complex,stella2025cognitive,amancio2012using,millan2025topology}.

A substantial body of research has investigated scientific impact prediction through diverse methodological lenses. Traditional bibliometric approaches frequently leverage early citation counts and journal-level indicators to estimate long-term influence~\cite{DBLP:journals/corr/WangSB13,STEGEHUIS2015642,ABRAMO201932}, though some studies caution that cumulative citation measures may overstate predictability due to intrinsic self-correlation~\cite{penner2013predictability}. Parallel to these metrics, advances in Natural Language Processing (exemplified by pretrained models like BERT \cite{DBLP:journals/corr/abs-1810-04805} and its domain-specific variant SciBERT \cite{beltagy-etal-2019-scibert}) demonstrate that scientific text can be effectively encoded for downstream tasks. Building on this, previous work evaluated a broad spectrum of text embeddings, including OpenAI's ADA-002, achieving robust performance in top-paper classification across multiple temporal horizons~\cite{vital2024predictingcitationimpactresearch}.

Complementing textual analysis, graph-based methods have gained prominence. DeepCCP \cite{zhao2022utilizing} established that citation-network structure alone offers competitive predictive power, while H2CGL \cite{he2023h2cgl} integrated the temporal evolution of citation contexts via graph neural networks. Bridging these domains, SPECTER \cite{cohan-etal-2020-specter} pioneered the fusion of textual and structural data by training document-level embeddings on citation signals. Most recently, the focus has shifted toward Large Language Models (LLMs). For instance, \citep{azad2025predicting} introduced retrieval-augmented generation for impact forecasting, while \citep{ye2026large} explored whether LLMs could identify highly cited papers from text and metadata alone, albeit highlighting risks of noisy or unhelpful context.

Despite these advances, a controlled comparison of these paradigms remains elusive. Extant studies frequently diverge in their target definitions, input signals, temporal assumptions, and susceptibility to data leakage. Notably, many traditional models rely on early citation counts or journal-level metrics -- indicators that are often unavailable or conceptually misaligned with strict cold-start scenarios. Furthermore, while LLM- and RAG-based approaches show promise, the conditions under which retrieved context provides a genuine signal versus introducing noise or ambiguity remain poorly understood. This underscores a critical research gap: under a unified, temporally consistent protocol, how do bibliometric, textual, graph-based, and LLM-based methods truly compare? Moreover, what is the marginal contribution of graph-retrieved peer context beyond the intrinsic information provided by the target paper itself?

The main objective of this study is to investigate how diverse information sources and inference paradigms contribute to the early prediction of scientific impact within a unified, temporally consistent evaluation framework. Specifically, we pursue three goals: (i) to assess the predictive value of citation-based and textual-similarity graph representations; (ii) to compare structural and hybrid structural-semantic embeddings derived from these graphs; and (iii) to determine whether Large Language Model (LLM) inference benefits from graph-retrieved contextual neighbors beyond the information intrinsic to the target paper. To achieve this, we examine impact prediction through two complementary lenses. First, we implement a graph-based supervised pipeline where citation and textual-similarity networks facilitate the construction of structural and hybrid representations. Second, we reformulate the task as an LLM-based probabilistic inference problem, evaluating target papers both in isolation and alongside graph-retrieved neighbors. In both settings, the goal is to predict whether a paper will rank within the top $P\%$ of its journal cohort $Y$ years post-publication.

This study extends prior research on text-based impact prediction by explicitly incorporating relational structure and evaluating a fundamentally different inference paradigm based on LLMs. Rather than comparing disparate methods across inconsistent datasets or temporal assumptions, we evaluate graph embeddings, hybrid representations, and GraphRAG-style prompting under identical constraints. 
This unified design enables a direct assessment of the respective roles played by citation structure, semantic similarity, and retrieved peer context.

Our contributions are fourfold: (1) we introduce a temporally consistent framework for top-paper prediction based on cohort-normalized impact across multiple horizons; (2) we compare citation and textual-similarity graphs across various directionality and weighting configurations; (3) we formalize the prediction task for LLM-based inference, testing the marginal utility of graph-retrieved context; and (4) we examine the robustness of these patterns across multiple journals to assess how graph and LLM-based strategies behave in diverse publication environments.

Our study yields several key insights. In the graph-based setting, our main results indicate that graph design choices significantly influence predictive performance. Among these, edge directionality (which captures the asymmetric structure of citations) emerges as the most critical factor, whereas edge weighting based on textual similarity contributes only marginal improvements. The optimal configuration combines structural embeddings from Node2Vec with textual embeddings derived from OpenAI models. This setup achieves an AUC of 0.84–0.85 and exhibits a rapid performance increase within the first two years post-publication.
In contrast, the GraphRAG-based experiments reveal a markedly different trend. While all configurations achieve strong performance (with peak AUC values ranging from approximately 0.85 to 0.87 in later horizons) the most notable finding is that omitting graph-retrieved context frequently improves results. Target-only LLM inference consistently achieves a more rapid ramp-up and yields equal or superior accuracy compared to context-augmented variants. This counterintuitive pattern persists across multiple journals (e.g., Journal of Informetrics, PNAS, and PRL), suggesting that the additional contextual information introduces redundancy or noise rather than a complementary signal.

\section{Methodology}
\label{chapter:methodology}

This section outlines the methodology adopted in this study, organized into two complementary sections. First, we detail the \emph{graph-based prediction} pipeline (Section \ref{chapter:graphmethod}), in which citation and textual-similarity graphs are constructed, mapped to structural and hybrid embeddings, and used as inputs for supervised ``top-paper'' prediction. Next, we introduce the \emph{LLM-based GraphRAG} framework (Section \ref{chapter:graphragmethod}), where the same graph structures are repurposed as retrieval mechanisms to provide contextual evidence for a large language model. Together, these components facilitate a comparison between compact vector representations and retrieval-augmented reasoning, while maintaining consistent target definitions and temporal evaluation protocols across both settings.

\subsection{Graph-based paper impact prediction}
\label{chapter:graphmethod}

The adopted methodology aims to predict whether a paper will rank among the top $P\%$ most cited papers within its {peer cohort} (i.e., papers in the same number of years after its publication and in the same journal) after a prediction horizon of $Y$ years following publication. 
To achieve this, we combine (i) \emph{structural information} derived from complex networks (citation and similarity graphs) and (ii) \emph{semantic information} derived from the paper's abstract via LLM embeddings. These information sources are used either independently or jointly as input features to a neural network classifier. We evaluate multiple experimental scenarios to assess predictability, robustness, and performance as a function of time since publication.
A high-level summary of the pipeline is shown in Figure \ref{fig:methodology}. Conceptually, the full method can be decomposed into four phases:

\begin{figure}[h]
\centering
\includegraphics[angle=-90, scale=0.6]{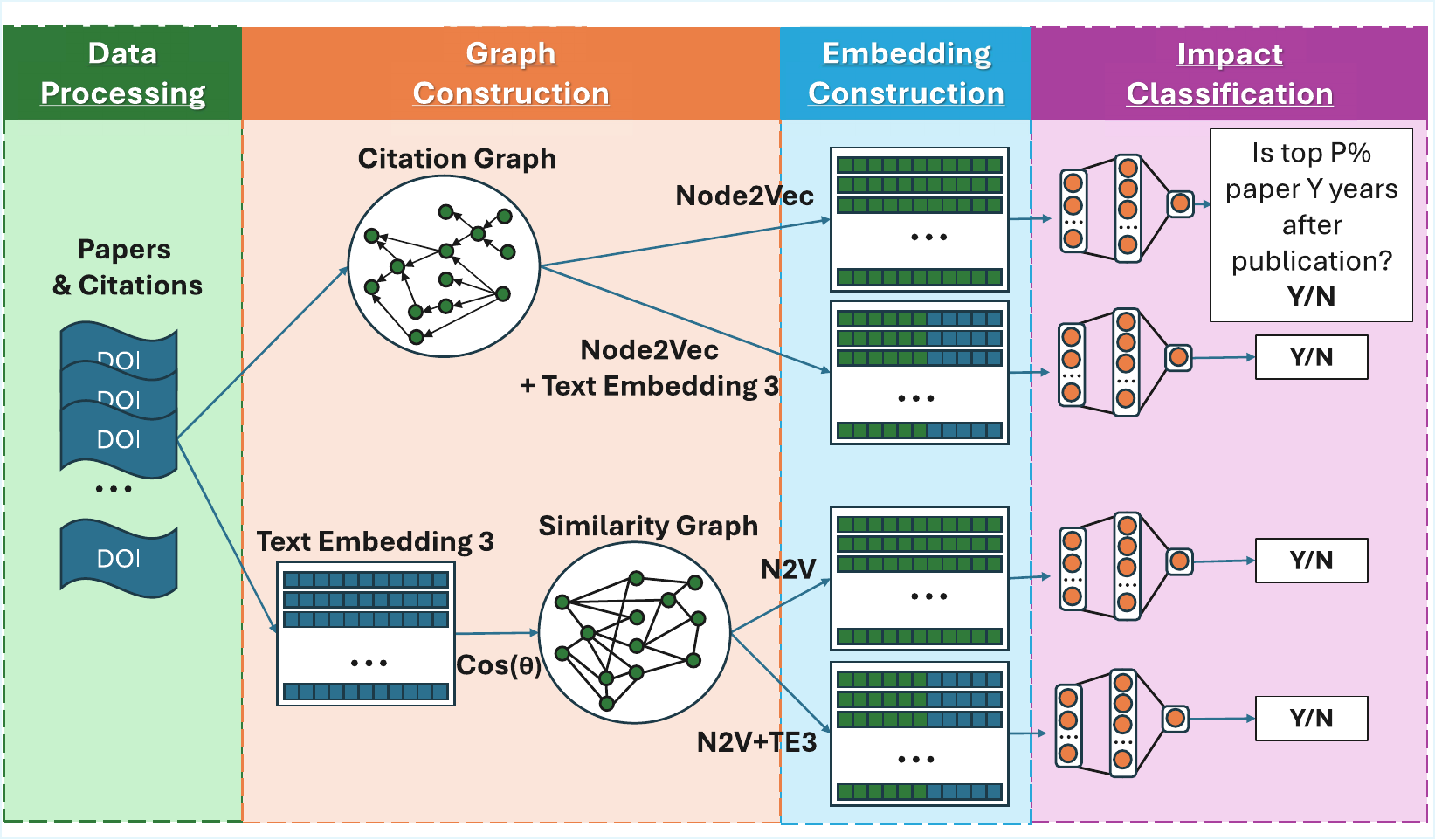}
\caption{{Overview of the methodology employed to predict scientific impact.} The process begins with data collection from scientific papers and citations, followed by the construction of two graphs: a citation graph and a textual similarity graph. Node embeddings are generated via Node2Vec alone or concatenated with OpenAI's textual based abstract embeddings. These are then used as inputs to neural classifiers tasked with predicting whether a paper will rank among the top $P\%$ in citations after $Y$ years of publication.}
\label{fig:methodology}
\end{figure}

\begin{enumerate}

    \item \emph{Data processing:} we collect the paper dataset and the historical yearly citation trajectories for each paper, and define targets for each year after publication across multiple horizons and impact thresholds.
    
    \item \emph{Graph construction:} we construct a citation network based on references among papers within the same journal and develop a similarity graph derived from the cosine similarity of OpenAI’s text-based abstract embeddings.
    
    \item \emph{Embedding construction:} we use Node2Vec~\cite{DBLP:journals/corr/GroverL16} to generate structural embeddings for both citation and similarity graphs, which are duplicated and concatenated with the previously computed GPT embeddings from the abstracts.
    
    \item \emph{Impact classification:} the different embedding representations are used as inputs to a neural network classifier to predict whether a paper will perform among the top papers after $Y$ years. The prediction is evaluated under the accumulated number of citations until the reference year.
    
\end{enumerate}

\subsubsection{Data processing}

\begin{figure}[h]
\centering
\includegraphics[scale=.75]{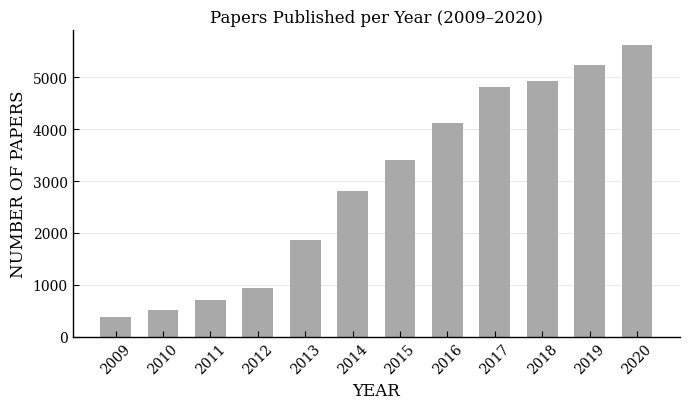}
\caption{{Number of papers published annually in ACS Applied Materials \& Interfaces} (American Chemical Society, ACS).}
\label{fig:dataset}
\end{figure}

The first part of this work started by selecting $35{,}354$ academic papers from the journal \emph{ACS Applied Materials \& Interfaces} (American Chemical Society, ACS), spanning 11 publication years from 2009 to 2020 and exhibiting a natural year-over-year growth in volume. The number of papers published yearly is shown in Figure \ref{fig:dataset}. The only information used at this stage was the Digital Object Identifier (DOI), year of publication, title, and abstract. This study extends the approach proposed in~\cite{vital2024predictingcitationimpactresearch}, where the same dataset was used with traditional textual embeddings and machine learning models for scientific impact prediction. As in \cite{vital2024predictingcitationimpactresearch}, we observed a stable distribution of highly cited papers across topics, preserving a diverse set of inputs for assessing the proposed models. 

After selecting the papers, we used an external API to collect the yearly citation counts of each article after publication. Because citation opportunity is strongly shaped by publication year, raw citation counts are not directly comparable across papers published in different years. This is due not only to differences in exposure time, but also to the well-known year-over-year inflation in citations driven by the growth of scientific production, indexing coverage, and referencing behavior~\cite{petersen2019inflation}. To account for this, we adopt a two-level temporal normalization strategy: (i) citations are analyzed by \emph{years since publication}, and (ii) impact labels are defined using \emph{within-cohort percentiles}. In practice, this means that each paper is compared only with papers from the same journal at the same time after publication, which enables fairer cohort-level comparisons and mitigates temporal bias.

Based on this framework, we quantify impact using the \emph{Accumulated Citation Count (ACC)} for each horizon $Y \in \{0,\dots,10\}$. For a paper $i$, the accumulated citations at horizon $Y$ are defined as $\mathrm{ACC}_{i,Y} = \sum_{k=0}^{Y} C_{i,k}$, where $C_{i,k}$ is the number of citations received in year $k$ after publication. Papers are then ranked within each cohort according to $\mathrm{ACC}_{i,Y}$, and a binary label is assigned indicating whether the paper belongs to the top $P\%$ of its cohort at that horizon. By construction, $\mathrm{ACC}_{i,Y}$ is non-decreasing with $Y$.

The prediction target is therefore binary: a paper receives label 1 if it is a top paper under the chosen percentile threshold, and 0 otherwise. We evaluate thresholds $P \in \{10,20,30,40,50\}$ and horizons $Y \in \{0,\dots,10\}$ whenever enough years are observable in the cohort. For each horizon, we estimate the empirical citation distribution within the cohort, compute the corresponding percentile cutoff, and assign labels accordingly. This produces a family of related prediction tasks that jointly capture short-, medium-, and long-term scientific impact.

\subsubsection{Graph Construction}

Once the database was organized into $Y$-year post-publication windows and the target metric labels were defined, we used the data available up to that point to construct two graphs representing complementary relational views: citations and semantic similarity. For both graph types we created four variations based on (i) edge direction (directed vs.\ undirected) and (ii) edge weights (unweighted vs.\ weighted). In the unweighted setting, all edges have weight 1. In the weighted setting, edge weights are given by cosine similarity between the textual embeddings of the two abstracts (defined in the similarity graph section, and also applied as an experimental variant in the citation graph). We evaluate weighted graph variants because previous results indicate that random-walk-based methods can recover relevant aspects of link-weight structure when weight information is incorporated into the walking dynamics \cite{vital2025recovering}. The integration of citation links and similarity-based connections is further supported by evidence that local similarity patterns and machine learning models offer complementary predictive insights in Science of Science research \cite{vital2022comparative}.

A central design principle when constructing the graphs is \emph{temporal consistency}: edges must never connect a paper to a ``future'' paper. This rule is enforced for both the citation-based and similarity-based graphs, thereby mitigating temporal leakage in the representation stage.

To construct the \emph{citation graph}, we restricted our scope to articles published within the same journal between 2009 and 2020. This boundary was intentionally set to maintain scalability; incorporating papers outside this scope would result in a prohibitively large graph, which is unrepresentative of practical scenarios where researchers often work with restricted or partial corpora. Within this defined set, edges represent observed citation relationships.

While citations are inherently directed (reflecting the specific act of referencing) we evaluated both directed and undirected variants to quantify the model's sensitivity to directionality. Furthermore, although citation edges are naturally binary, we introduced a weighted variant where edge weights correspond to the cosine similarity between abstract embeddings. This approach allows us to incorporate topical proximity as a proxy for edge strength within the structural citation graph.

To construct the \emph{similarity graph}, we leveraged the OpenAI \texttt{text-embedding-3-large} model  to represent each abstract as a 3,072-dimensional vector~\cite{openai_embeddings_v3_2024}. Subsequently, we performed an all-pairs similarity computation to determine the cosine similarity between all abstracts. For each paper, edges were established by selecting the top $K$ most similar papers, where $K \in \{3, 5, 7, 9\}$.

Mirroring the citation graph construction, we evaluate both directed and undirected variants. In the directed similarity graph, edges originate from the source paper and point toward its top-$K$ most similar preceding papers. In the undirected variant, we omit edge orientation, treating similarity as a symmetric relationship for representation learning. Furthermore, we compare weighted versus unweighted configurations: the former uses cosine similarity as the edge weight, whereas the latter assigns a constant weight of 1 to all edges.

\subsubsection{Embedding construction}

After the graph construction phase, we have the following graph families: (i) the citation graph built from citation relations, and (ii) the similarity graph built by connecting top-$K$ most similar papers, producing four similarity-graph variants according to $K \in \{3,5,7,9\}$. Each graph family is further expanded by two edge direction types (directed vs.\ undirected) and two weighting strategies (unweighted vs.\ weighted). In the weighted setting, the weight is the cosine similarity between the abstract embedding vectors of the two endpoints; this applies naturally to similarity graphs and is also evaluated as an experimental proxy in citation graphs.

To construct the feature sets for our primary downstream task (paper impact classification) we employ a method to extract structural graph information into a numerical format compatible with machine learning models. Specifically, we use Node2Vec~\cite{DBLP:journals/corr/GroverL16}, a random-walk-based framework that samples node sequences via a stochastic mechanism with tunable exploration behavior. Following sequence collection, Node2Vec optimizes a skip-gram objective \cite{41224, DBLP:journals/corr/MikolovSCCD13} to learn distributed node representations. The resulting embedding vectors encapsulate local neighborhood context and broader structural regularities, providing a robust foundation for node-level classification tasks (including the impact classification problem studied here).

While other graph-embedding methods could also be used, we adopted {Node2Vec} because it offers a well-established and flexible random-walk-based framework for encoding structural information through sampled node sequences. Prior work also suggests that embeddings derived from different walk biases often yield relatively stable downstream performance, indicating that sequence-based methods can robustly recover meaningful structural patterns even under distinct traversal heuristics. We did not consider other alternatives such as ProbWalk, ARGEW, or GNN-based models~\cite{10.1371/journal.pone.0312863,wu2021probwalk,kim2023node,ZHOU2024121072, HE2023222}, since the goal of this study was not to compare embedding architectures, but to isolate the effects of graph construction, edge configuration, and the combination of structural and textual information within a single consistent embedding framework.

We maintained consistent Node2Vec hyperparameters across all experiments, employing an embedding dimension of 256, a walk length of 120, and 20 walks per node. The context window size was set to 10, while the return ($p$) and in-out ($q$) parameters were both fixed at 1.0, effectively resulting in a standard unbiased random walk.

We then define two alternative input representations for the classifier:

\begin{enumerate}

\item \emph{Structural embedding:}  in this configuration, we exclusively leverage the structural information of the graphs, encoded as Node2Vec embedding vectors -- a strategy henceforth denoted as N2V. Prioritizing structural data is particularly pertinent for the citation graph, where the fundamental signal resides in the citation link $A \rightarrow B$; here, graph embeddings represent a well-established paradigm for capturing such topological patterns. This representation bypasses direct textual features, despite their inherent richness. Within the similarity graph, while text informs the underlying structure, the application of graph embeddings serves to superimpose structural organization onto a semantic topology. Although this introduces a degree of redundancy, the approach remains advantageous, as Node2Vec elevates simple pairwise similarities into higher-order neighborhood representations.

\item \emph{Structural and textual embedding}: in this configuration, we concatenate the structural vectors generated by Node2Vec with the corresponding OpenAI text embeddings~\cite{openai_embeddings_v3_2024} previously derived from the abstracts. This integrated representation is henceforth referred to as N2V+TE3. This concatenation is particularly vital for the citation graph, as it introduces textual context into a model that would otherwise rely exclusively on topological structure. For the similarity graph, while this process might seem partially redundant (given that the graph is itself constructed from textual embeddings) the two components provide distinct signals: one is a graph-transformed representation that captures higher-order relational neighborhoods, while the other provides the raw semantic context of the abstract. Empirically, combining these complementary views can enhance predictive performance, aligning with prior findings that citation patterns and content similarity are not perfectly correlated and often provide additive information~\cite{amancio2012using}.
        
\end{enumerate}

\subsubsection{Impact classification}

In the final phase, we use the embeddings from the previous phase as inputs to a supervised classification model that predicts whether each paper will be a ``top paper'' under a given definition, percentile threshold $P \in \{10,20,30,40,50\}$, and prediction horizon $Y \in \{0,\dots,10\}$ (when observable). Concretely, each training instance corresponds to a paper, represented either by its N2V embedding alone or by the concatenated N2V+TE3 vector, and labeled according to the chosen target.

We adopt a feed-forward neural network classifier with two dense hidden layers of 64 and 32 neurons, respectively, and a sigmoid output neuron for binary prediction. Training is performed with batch size 2048, for up to 80 epochs, using a validation split of 15\% and early stopping with patience of 5 epochs. For efficiency in I/O and preprocessing, a chunksize of 200{,}000 is used during dataset handling.
To mitigate class imbalance and ensure robust estimates, we construct balanced datasets for each $(Y,P,\text{target})$ configuration. Specifically, the positive class consists of papers in the top $P\%$ of their cohort, and we randomly sample an equal number of negative examples from the remaining papers. For example, for $Y=1$ and $P=20\%$, the dataset contains 7{,}071 positive papers, and we sample 7{,}071 negatives from the remaining 80\%. This balanced sampling procedure is repeated 10 times with independent random sampling (i.e., 10 repeated runs), and we report the average classification performance across repetitions.

Model evaluation is conducted using the Area Under the Receiver Operating Characteristic Curve (AUC-ROC). This metric is computed for each combination of (graph type, $K$ for similarity graphs, directionality, weighting, embedding representation, target definition, percentile threshold, and horizon), enabling a comprehensive factorial analysis of how structural and semantic signals contribute to impact predictability over time. Additionally, temporal leakage is mitigated at the representation level by enforcing temporally feasible edges in both graph families (no connections to future papers), aligning graph construction with the causal availability of information at publication time.

\subsection{LLM-based paper impact prediction with GraphRAG}
\label{chapter:graphragmethod}

\begin{figure}[h]
\centering
\includegraphics[angle=-90, scale=0.60]{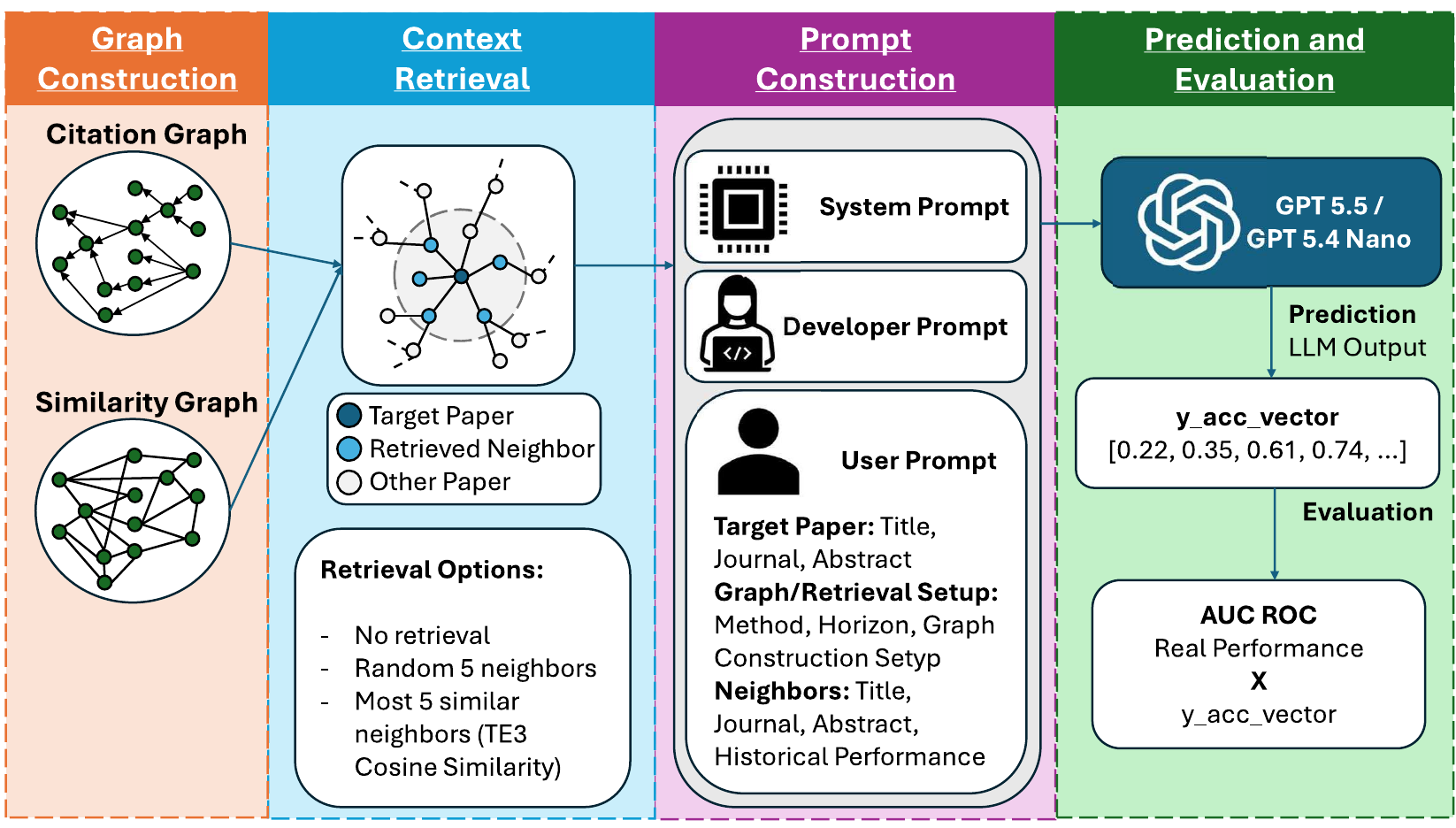}
\caption{{Overview of the LLM-based GraphRAG methodology for top-paper prediction.} 
The process reuses the citation and textual-similarity graphs constructed in the graph-based stage, but replaces fixed graph embeddings and neural classification with a retrieval-augmented LLM inference pipeline. For each target paper, a graph-based context is retrieved according to one of three settings: no retrieval, random neighbors, or the top five most similar neighbors based on cosine similarity of OpenAI's \emph{Text Embedding 3} abstract representations. The retrieved information is then incorporated into a structured prompt composed of system, developer, and XML-based user instructions, including the target paper metadata, graph/retrieval setup, neighbor information, and historical performance vectors. GPT-5.5 (or GPT 5.4 Nano) produces a multi-horizon accumulated-citation probability vector, \texttt{y\_acc\_vector}, which is evaluated against the real top-$20\%$ labels for each prediction horizon using AUC-ROC.}
\label{fig:methodologygraphrag}
\end{figure}

In the second phase of this study, we reformulate the impact prediction problem as a retrieval-augmented probabilistic inference task executed by a Large Language Model (LLM). While the predictive objective remains identical (identifying whether a target paper will rank among the top $20\%$ most-cited papers within its journal cohort after a $Y$-year horizon) the underlying mechanism shifts. We maintain the exact target definitions, cohort-based normalization, and temporal evaluation protocols established in the graph-based methodology; these are omitted here for brevity. The fundamental distinction lies in the processing of structural and semantic data: rather than compressing these features into static vectors for a neural classifier, we utilize the graph as a retrieval mechanism. This allows the LLM to perform direct reasoning over the target paper and, where applicable, a selective set of graph-retrieved neighbors. Consequently, the graph transitions from a source for node embeddings to a substrate for GraphRAG-style context selection.
The pipeline is summarized in Figure \ref{fig:methodologygraphrag} and is structured into four distinct phases:

\begin{enumerate}

    \item \emph{Graph construction:} we leverage the citation and textual-similarity graphs established in the previous phase, adhering to the same temporal-consistency constraints. In the main \textcolor{red}{ACS} experiments, we evaluate multiple graph variants and sample approximately 10\% of the nodes for LLM inference. Cross-journal evaluations conducted on \emph{Informetrics}, \emph{PNAS}, and \emph{PRL} utilize a directed, unweighted configuration.

    \item \emph{Context retrieval:}  for each sampled target paper, we extract a small graph neighborhood to serve as contextual information for the prompt. We evaluate two retrieval strategies: (i) randomly selected neighbors and (ii) the top-5 most similar neighbors, determined by the cosine similarity of abstract embeddings generated via the OpenAI \emph{text-embedding-3} model. Additionally, we include a null baseline that excludes retrieved neighbors to assess the impact of graph context.
    
    \item \emph{Prompt construction:} We use GPT-5.5 or GPT 5.4 Nano via a three-tier prompting protocol comprising a \emph{system prompt}, a \emph{developer prompt}, and an XML-structured \emph{user prompt}. This framework encapsulates the graph and retrieval configurations, the target paper’s metadata, and -- where applicable -- the retrieved neighbors alongside their historical performance vectors.

    \item \emph{Prediction and evaluation:} for each target paper, the LLM generates a multi-horizon probability profile in a single inference pass. In this study, we focus specifically on the cumulative citation prediction output (\texttt{y\_acc\_vector}), evaluating it across all time horizons against the binary labels established in the graph-based phase. Performance is quantified using the AUC-ROC curve.

\end{enumerate}

\subsubsection{Graph construction}

The LLM-based experiments use the graphs constructed in the previous stage; consequently, no additional graph-construction procedures are introduced here. We retain the same two graph families: (i) citation graphs and (ii) textual-similarity graphs. The temporal-consistency rule remains strictly enforced, ensuring no paper connects to future publications. While this is inherently maintained in citation graphs by the direction of the citations, it is explicitly applied to textual-similarity graphs. By restricting connections to semantically similar papers available at the time of publication, we effectively prevent temporal leakage during retrieval.

In the graph-based experiments, we evaluated multiple graph variants by combining directionality (directed vs. undirected) and weighting (weighted vs. unweighted), mirroring the configurations used in the graph-based setting. From each graph, we randomly sampled approximately 10\% of the nodes for LLM inference, representing 3400 papers out of a total of 35,354. While this represents a small fraction of the network, it yields a robust evaluation set; each sampled paper generates predictions across multiple temporal horizons (up to 10 years post-publication), significantly increasing the number of effective prediction instances. This subsampling approach ensures a representative analysis of the method’s performance while maintaining computational feasibility, particularly as prompt-based inference is considerably more resource-intensive than neural classification using fixed embeddings.

In order to directly compare prompting with and without retrieved neighbors, we fixed a single graph configuration (directed and unweighted) to isolate the impact of contextual retrieval from the effects of graph topology. This simplified setting was also applied to our cross-journal experiments. To evaluate generalization, we replicated the LLM-based protocol following the methodology of \cite{vital2024predictingcitationimpactresearch} across three additional journals: Informetrics, PNAS, and PRL. Specifically, we sampled 400 nodes from Informetrics ($N=391$), 2050 nodes from PNAS ($N=20{,}398$), and 1650 nodes from PRL ($N=16{,}439$). Notably, unlike the graph-based neural experiments, this stage does not employ Node2Vec, as the graph structure is used exclusively for retrieval rather than embedding generation.

\subsubsection{Context retrieval}

For each sampled target paper, we extracted a local neighborhood from the graph to serve as contextual information within the prompt. We evaluated two distinct retrieval strategies: (i) random sampling from the target node’s immediate graph neighbors, and (ii) similarity-based selection, where we identified the top five most similar papers based on the cosine similarity of their Text Embedding 3 (OpenAI) abstract representations. In both scenarios, a maximum of five neighbors was included. This design facilitates a direct comparison between semantically guided retrieval and a stochastic graph-based baseline. Furthermore, it enables us to assess whether the LLM benefits from explicitly receiving papers that are close to the target either structurally or semantically.

Beyond the GraphRAG configurations, we established a null (context-free) baseline wherein the LLM receives only the target paper’s information without any neighborhood context. This baseline serves as a critical control to determine whether graph-retrieved context yields a measurable improvement in prediction or if additional context introduces noise. To maintain temporal integrity, whenever neighbor citation trajectories were included in the prompt, we only disclosed information available up to the target paper's publication year. This restriction ensures temporal realism, preventing the LLM from indirectly accessing future evidence through the neighbors' historical data.

\subsubsection{Prompt construction}

We employed GPT-5.5 and GPT 5.4 Nano as the underlying models, configuring the prompting protocol to function as a specialized scientific impact prediction engine rather than a general-purpose assistant. To achieve this, each request was structured into three distinct layers: a system prompt, a developer prompt, and a programmatically generated user prompt. Beyond improving conceptual modularity, this architecture was adopted to optimize computational efficiency; by isolating static instructional components from instance-specific data, we maximized the potential for prompt caching, thereby significantly reducing experimental costs. The full schema is detailed in Appendix \ref{appendix:prompt_templates}.

The system prompt enforces strict adherence to a JSON-only output format, while the developer prompt formalizes the predictive task: estimating the probability that a target paper will reach top paper status under ACC criteria for each horizon year.

The user prompt is serialized in XML format, ensuring a clear and machine-readable structure for the model. At a minimum, each prompt contains a \texttt{<CONFIG>} block and a \texttt{<TARGET>} block. The former specifies experimental parameters -- including graph identifiers, retrieval strategy, directionality, weighting, and quantile thresholds -- while the latter encapsulates the target paper’s DOI, metadata, abstract, and temporal parameters (horizon and maximum observable years). In the GraphRAG setting, a \texttt{<NEIGHBORS>} section is included, containing serialized metadata and historical performance vectors for each retrieved paper. For the context-free baseline, this neighborhood block is entirely omitted. This structured serialization ensures the protocol remains explicit, reproducible, and robust across different graph topologies and journal datasets

\subsubsection{Prediction and evaluation}

For each target paper, the LLM produces a single structured response centered on a probability vector for top-paper prediction across all requested horizon years, together with additional auxiliary outputs included for completeness. Thus, unlike the graph-based classifier, which solves separate binary classification problems for each $(P,Y,\text{target})$ configuration, the LLM generates a full multi-horizon prediction profile for a target paper in a single inference.

In the analyses reported in this work, we only use the probability output associated with accumulated-citation prediction, \texttt{y\_acc\_vector}. Accordingly, the LLM probabilities derived from \texttt{y\_acc\_vector} are matched against the same binary ground-truth labels defined in the graph-based stage, and performance is evaluated horizon by horizon using only AUC-ROC.

Taken together, this methodology allows us to evaluate not only whether graph-retrieved context helps LLM-based impact prediction, but also whether such retrieval remains useful when compared against a simpler prompt containing only the target paper. By repeating the same design across multiple graph configurations and across additional journals, we can assess both the robustness of GraphRAG-style retrieval and its actual contribution to predictive performance.

\section{Results and discussion}
\label{chapter:results}

This section details the empirical findings of the proposed methodologies and explores their implications for predicting scientific impact. Section \ref{chapter:graphresults} initiates the analysis by examining graph-based models, specifically assessing how graph construction, representation learning, and label formalization influence predictive performance. Subsequently, Section \ref{chapter:graphrragesults} evaluates the LLM-based GraphRAG framework, highlighting the influence of retrieved context, varied prompting strategies, and generalization across diverse journals. Collectively, these results offer a comparative perspective on impact prediction using both structured graph representations and retrieval-augmented language model inference.

\subsection{Graph-based prediction results}
\label{chapter:graphresults}

This subsection presents the primary results of the graph-based prediction pipeline through three complementary lenses. Section \ref{chapter:graphcompcitvstext} initiates the discussion by evaluating the performance of citation-based versus textual-similarity graphs, while also assessing the relative merits of purely structural and hybrid structural-textual representations. In Section \ref{chapter:graphsensitopktext}, we investigate the sensitivity of the textual-similarity approach to the top-$K$ neighborhood parameter selected during graph construction. Finally, Section \ref{chapter:graphquantilecomp} examines how predictive difficulty fluctuates as the definition of 'top papers' is modulated through increasingly selective quantile thresholds. 

\subsubsection{Comparison between citation and textual-similarity graphs}
\label{chapter:graphcompcitvstext}

\begin{figure}[h]
\centering
\includegraphics[scale=0.5]{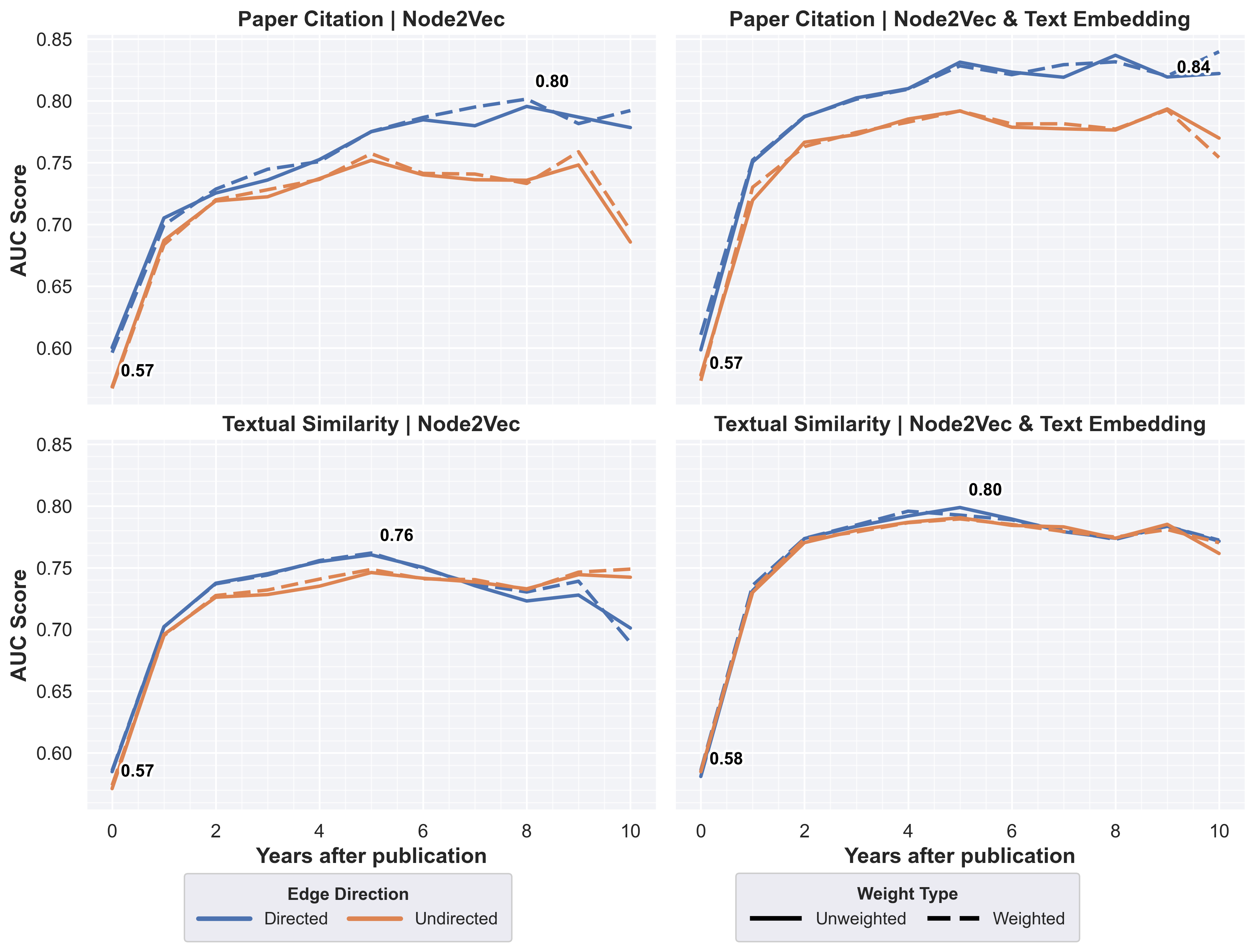}
\caption{{Comparison of graph construction strategies and input representations for top-20\% paper prediction over time.} The figure reports AUC scores across years after publication for four configurations: two graph construction strategies, \emph{Paper Citation} and \emph{Textual Similarity}, and two neural-network input types, Node2Vec alone and Node2Vec concatenated with OpenAI's textual embedding. Colors indicate edge directionality (blue: directed, orange: undirected), while line style indicates whether edge weights are used (solid: unweighted, dashed: weighted).}
\label{fig:GraphNNclassification}
\end{figure}

Figure~\ref{fig:GraphNNclassification} presents the AUC scores for the classification of papers belonging to the top 20\% of their respective cohorts for each year following publication. We evaluate two distinct graph construction strategies -- \emph{Paper Citation} and \emph{Textual Similarity} -- against two input representations for the neural classifier: Node2Vec alone, to capture purely structural features, and a hybrid approach concatenating Node2Vec with OpenAI’s \emph{Text Embedding 3} to integrate both structural and textual information. 

A consistent pattern emerges across all four panels: predictive performance begins at a modest yet above-random level (AUC $\approx$ 0.57--0.60) at year 0, followed by a sharp increase during the first two to three years post-publication, typically reaching the 0.70--0.77 range. 
After this initial growth phase, the curves either improve more incrementally or reach a stable plateau between years 3 and 8, followed by minor oscillations and, occasionally, a slight decline in the later years. This temporal trajectory aligns with previous findings~\cite{vital2024predictingcitationimpactresearch}, where textual embeddings were applied directly to impact prediction. From a practical standpoint, the rapid early gain suggests that high-impact papers harbor detectable signals shortly after release. At year 0, however, performance is constrained by a 'cold-start' effect; while a paper may contain latent predictive precursors, it lacks sufficient time for dissemination and the accumulation of a measurable citation history. Once this threshold is surpassed, the challenge shifts to maintaining performance over medium- and long-term horizons -- a regime where differences between modeling configurations become most pronounced.

The citation-based graph emerges as the superior configuration across all tests. Within the \emph{Paper Citation} framework, the highest performance is achieved using directed graphs, particularly when structural and textual features are integrated. While Node2Vec alone reaches a peak AUC of approximately 0.80, the addition of textual embeddings elevates this baseline to 0.84. The performance gap between directed and undirected variants is substantial, underscoring that edge directionality encodes critical predictive signals. In citation networks, direction does not merely represent a linkage; it preserves the chronological order and the causal flow inherent in the evolution of scientific knowledge. By discarding directionality, these temporal nuances are lost, leading to a concomitant drop in performance. This effect is significantly more pronounced in citation graphs than in similarity-based networks, where direction is often an imposed constraint for temporal consistency rather than an intrinsic semantic property.

By contrast, the inclusion of edge weights yields a comparatively marginal and inconsistent effect. In several instances, weighted and unweighted variants exhibit nearly identical performance, with unweighted versions occasionally outperforming their weighted counterparts across specific horizons. This finding carries significant practical weight: the computational overhead of calculating and maintaining similarity-based weights (particularly for large-scale corpora and high dimensional representations) yields diminishing returns. Our results suggest that edge weighting is not a primary driver of predictive accuracy for this task. This has direct implications for scalability and operational efficiency, indicating that unweighted architectures can be prioritized to reduce resource consumption without compromising model performance.

A significantly more consequential modeling choice is whether to include textual embeddings in the classifier input. In the citation graph, integrating textual embeddings yields a clear and practically meaningful gain, both in peak performance and the speed of convergence. For example, the Node2Vec-only configuration requires approximately four years post-publication to reach an AUC of 0.75; in contrast, the combined Node2Vec + Text Embedding setup achieves a comparable level within the first year. Similarly, while the former only approaches an AUC of 0.80 near year eight, the latter reaches this threshold by years three or four. Although the textual component adds a computational burden (introducing 3,072 additional input dimensions compared to the 256 used for Node2Vec) the substantial performance gains justify the cost when early, high-accuracy prediction is a priority.

Interestingly, the benefit of incorporating textual embeddings is significantly greater for citation graphs than for similarity graphs -- a result that is conceptually consistent. In similarity graphs, textual information is already inherent to the graph’s topology, as edges are defined by the cosine similarity between abstract embeddings. Consequently, reintroducing the same textual representation at the classifier stage may create redundancy. Conversely, in citation graphs, the structural representation is not directly derived from textual content; thus, the explicit addition of textual embeddings provides a genuinely complementary information source. Our empirical results confirm this interpretation: text serves as a powerful adjunct to citation structure, whereas its marginal utility is diminished when the graph is already constructed from semantic similarity.

For \emph{Textual Similarity} graphs, performance is more uniform across directionality and weighting configurations. As with citation networks, the best results emerge when combining structural and textual representations -- peaking at approximately 0.80 AUC, compared to roughly 0.76 for Node2Vec alone. However, unlike the citation case, the discrepancy between directed and undirected variants is marginal. This consistency aligns with the underlying graph semantics: while direction is intrinsically meaningful in citation networks, it primarily serves to impose temporal feasibility on the otherwise symmetric nature of semantic proximity in similarity graphs. The temporal profile also diverges; rather than a sustained upward trend, similarity-based graphs exhibit a sharp initial ramp-up followed by a plateau and a slight downward drift in later years. This suggests that while semantic similarity effectively identifies early and mid-term success signals, it is less robust than citation structure for long-horizon discrimination.

Taken together, these results reinforce a pivotal conceptual conclusion in our analyzed datasets: scientific impact is not a purely semantic phenomenon. Papers that share similar topics, terminology, or methodological frameworks do not necessarily converge toward similar citation outcomes. Instead, a paper's position within the evolving relational topology of scientific production is a critical determinant, and citation graphs provide a more direct representation of this structure. While semantics clearly enhances performance (often substantially) it serves as a complement rather than a substitute. Indeed, the highest-performing configuration is the one that integrates a citation-based relational backbone with explicit textual information. This observation has broader implications beyond impact prediction, particularly for the development of paper embeddings and GraphRAG systems: the most promising research direction lies not in replacing structure with text, but in explicitly modeling their interplay.

\subsubsection{Sensitivity analysis of Top-K textual similarity graphs}
\label{chapter:graphsensitopktext}

\begin{figure}[H]
\centering
\includegraphics[scale=0.48]{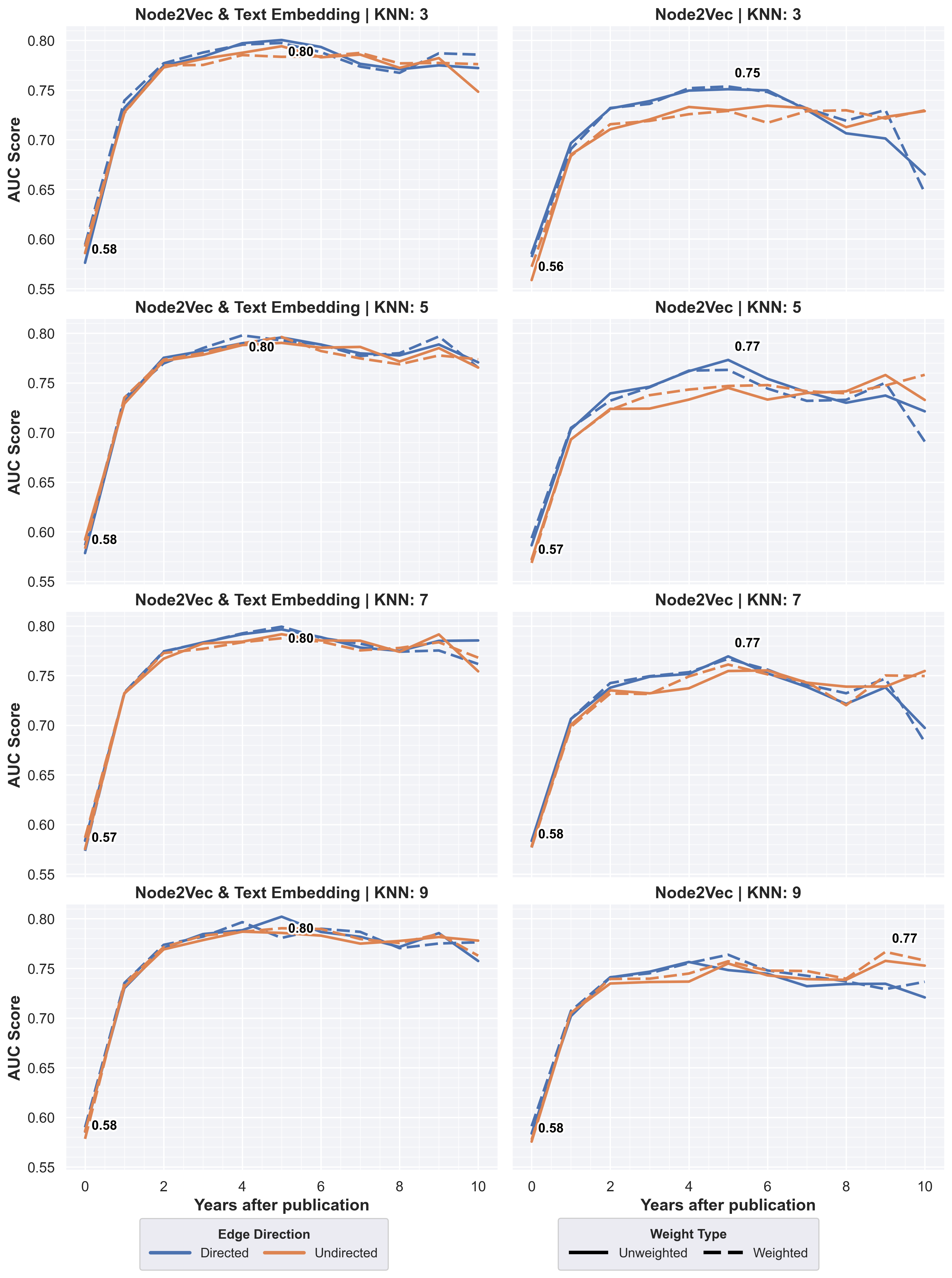}
\caption{{Performance comparison of textual-similarity graphs built with Top-$K$ neighborhoods.} The figure reports AUC scores across years after publication for textual-similarity graphs using four neighborhood sizes ($K=3,5,7,9$), with either Node2Vec alone or Node2Vec concatenated with textual embeddings. Colors indicate edge directionality, while line style indicates whether edge weights are used.}
\label{fig:GraphCompKNN}
\end{figure}

In Figure \ref{fig:GraphCompKNN}, we examine the textual-similarity graphs by varying the number of neighbors ($K$) from 3 to 9. The most prominent finding is that concatenating textual embeddings with Node2Vec consistently outperforms the standalone Node2Vec model across all values of $K$. This confirms that explicitly preserving the original textual embeddings at the classification stage remains beneficial, even when the underlying graph is already derived from textual similarity. Furthermore, the results demonstrate low sensitivity to the choice of $K$. For Node2Vec in isolation, increasing $K$ from 3 to 5 yields a marginal improvement, with the peak AUC rising from approximately 0.75 to 0.77; however, performance plateaus beyond $K=5$. In contrast, the combined \emph{Node2Vec + Text Embedding} representation exhibits even greater stability, as all tested values of $K$ produce nearly identical curves with peak AUCs near 0.80. These results suggest that when textual information is explicitly included in the input, the model's performance becomes remarkably robust to the neighborhood size used during similarity graph construction.

Another noteworthy pattern is that most configurations achieve peak performance between four and six years post-publication, followed by a plateau or a marginal decline. The primary exception is the standalone Node2Vec model with $K=9$, which reaches its maximum value later in the timeline. Nevertheless, this observation does not shift the general conclusion: the choice of representation exerts a more substantial influence than the specific value of $K$. From a practical perspective, $K=5$ emerges as an optimal compromise. It delivers peak or near-peak performance while maintaining greater efficiency (in terms of graph density, storage, and computational overhead) than larger neighborhoods. Consequently, as a rule of thumb, $K=5$ provides the most favorable balance between computational efficiency and predictive accuracy in the textual-similarity setting.

\subsubsection{Effect of quantile-based thresholds on top-paper prediction performance}
\label{chapter:graphquantilecomp}

\begin{figure}[h]
\centering
\includegraphics[scale=0.55]{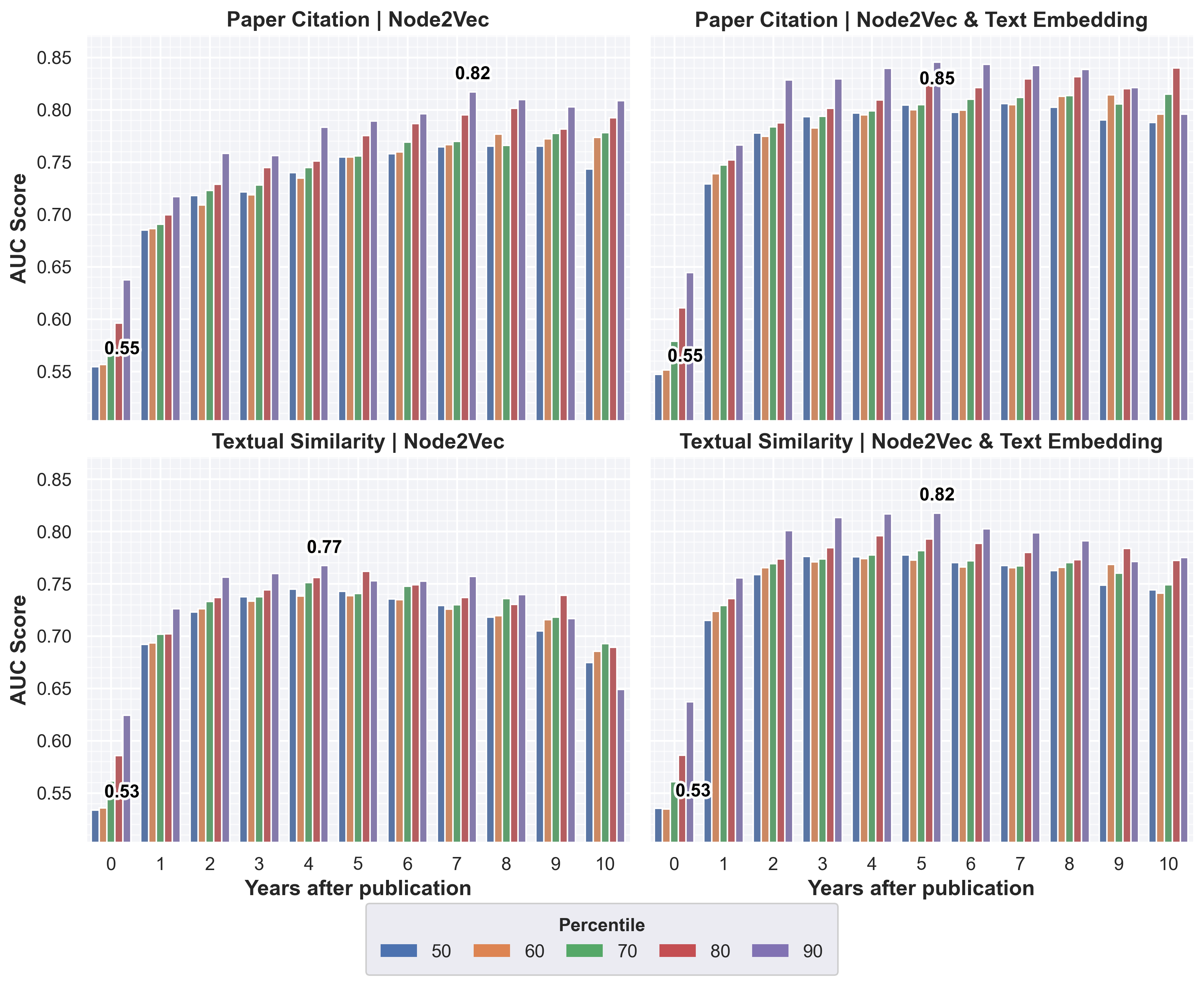}
\caption{{Effect of quantile-based thresholds on top-paper prediction performance.} The figure reports AUC scores across years after publication for four graph/input configurations for directed and weighted graphs: citation graph with Node2Vec, citation graph with Node2Vec concatenated with textual embeddings, textual-similarity graph with Node2Vec, and textual-similarity graph with Node2Vec concatenated with textual embeddings. Bars correspond to different percentile thresholds used to define the positive class (50, 60, 70, 80, and 90), where higher thresholds represent stricter top-paper criteria.}
\label{fig:GraphCompTOPK}
\end{figure}

In Figure \ref{fig:GraphCompTOPK}, we extend the experiment across various quantile thresholds ($50^{th}$ to $90^{th}$ percentiles in increments of 10) used to define the positive class. While the figure displays results exclusively for directed and weighted graphs, the observed trends were consistent across other graph variations. Overall, higher percentile thresholds systematically yield superior AUC values. This is a particularly insightful finding, as it suggests that ``exceptional'' papers (i.e., those significantly outperforming their cohort) are more readily identifiable than those that are merely above average. This pattern is especially pronounced up to approximately eight years post-publication, where more selective thresholds yield consistently better discrimination. 

The relative performance ranking among graph types and input representations remains consistent with our previous findings. Citation-based graphs continue to outperform textual-similarity counterparts, particularly when structural and textual features are integrated. Specifically, the addition of textual embeddings enhances citation graph performance from approximately 0.82 to 0.85, whereas the gain in textual-similarity graphs is more substantial, rising from 0.77 to 0.82. Peak performance typically occurs between years 4 and 7, suggesting that the distinction between future high-impact papers and the remainder of the cohort is most discernible during this period.

As observed in previous figures, there is a marked initial surge in predictive performance. While the baseline AUC at year 0 ranges between 0.53 and 0.55, numerous configurations surpass the 0.75 threshold within the first two years. This reinforces the notion that future high-impact papers often emit distinctive signals early in their life cycle. More broadly, these results indicate that the proposed methodology is particularly effective at identifying true outliers rather than distinguishing between moderate levels of success. From a practical standpoint, this is highly encouraging; many real-world applications (such as editorial screening, talent scouting, and portfolio prioritization) are primarily concerned with capturing papers most likely to achieve exceptional long-term impact.

\subsection{GraphRAG-based top-paper prediction}
\label{chapter:graphrragesults}

This subsection details the primary findings of our GraphRAG-based prediction pipeline through three complementary lenses. First, in Section \ref{chapter:graphrageffectneighborstrategy}, we evaluate alternative neighbor-retrieval strategies and investigate the performance of citation versus textual-similarity graphs when employed as contextual substrates for the LLM. Subsequently, in Section \ref{chapter:graphragpredictionperf}, we isolate the contribution of graph-augmented context by contrasting neighbor-enriched prompts against a context-free baseline. Finally, Section \ref{chapter:graphragcrossjournal} examines the transferability of these patterns across additional journals beyond the primary corpus. Collectively, these analyses elucidate the conditions under which graph-retrieved context enhances predictive performance and assess the robustness of the GraphRAG framework across diverse graph architectures and journal collections.

\subsubsection{Effect of neighbor retrieval strategy in GraphRAG-based prediction}
\label{chapter:graphrageffectneighborstrategy}

We employed the graph structure as a retrieval mechanism, selecting neighbors of the target paper and injecting them into the LLM prompt as contextual evidence. We evaluated multiple graph configurations by varying edge direction and edge weighting, and compared two neighborhood-selection strategies: (i) random neighbor retrieval and (ii) retrieval of the Top-5 most similar neighbors, where similarity was defined by cosine similarity between OpenAI \emph{Text Embedding 3} representations of the abstracts. Thus, unlike the graph-based neural approach, where the graph is primarily used to generate node embeddings, here the graph functions directly as the substrate for GraphRAG-style context augmentation.

For this experiment, we used a smaller and cheaper model, GPT-5.4 Nano, rather than the more expensive models used in the broader LLM evaluation. This choice was made to make the large number of graph-construction and retrieval-strategy combinations computationally and financially feasible. In our setup, GPT-5.4 Nano was approximately 27 times cheaper per execution than GPT-5.5 with low reasoning effort, with an average cost of about USD 0.00059 per execution compared with USD 0.016. Therefore, the results in this subsection should be interpreted not only as a test of GraphRAG retrieval, but also as an evaluation of how much signal a lightweight LLM can extract from graph-retrieved scientific context.

\begin{figure}[h]
\centering
\includegraphics[scale=0.5]{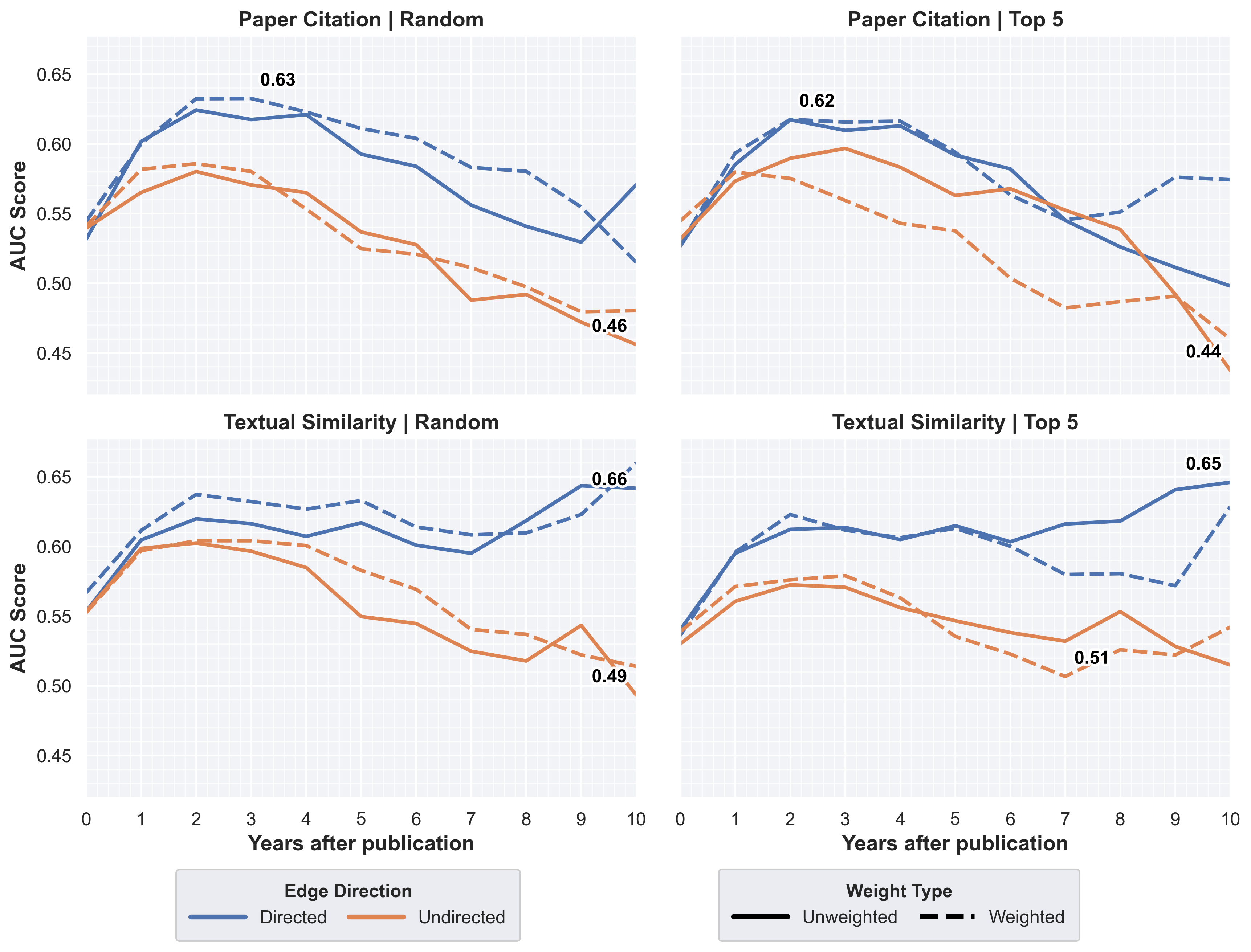}
\caption{{GraphRAG-based prediction performance under different neighbor retrieval strategies using GPT-5.4 Nano.} The figure reports AUC scores across years after publication for citation and textual-similarity graphs using two retrieval settings: random neighbors and Top-5 most similar neighbors. Results are shown across direction and weighting configurations on a randomly sampled subset of approximately 10\% of the network nodes.}
\label{fig:LLMCompSimCit}
\end{figure}

The results are shown in Figure \ref{fig:LLMCompSimCit}. Overall, the performance is substantially lower than what we observed in the graph-based neural models and in the stronger LLM experiments. Most configurations start around 0.53--0.56 AUC, which is slightly above random but still weak for practical prediction. The best results remain modest: the strongest citation-based configurations reach approximately 0.62--0.63 AUC, while the best textual-similarity configurations reach approximately 0.65--0.66 AUC. This indicates that, under this lightweight model, simply adding graph-retrieved neighbors to the prompt is not sufficient to produce strong top-paper prediction performance.

The most consistent pattern is the importance of edge direction. Directed graphs generally outperform undirected graphs across most panels and horizons, and the gap becomes especially visible in the later years. This is particularly clear in the textual-similarity graphs, where directed variants maintain or improve performance over time, while undirected variants tend to decline. This suggests that even in a GraphRAG setting, where the LLM reads textual context directly, the temporal and relational constraints encoded by direction still influence the quality of the retrieved evidence.

Edge weighting, in contrast, has a smaller and less consistent effect. Weighted and unweighted variants often behave similarly, and there is no clear evidence that weighting systematically improves prediction. This is important from a practical standpoint because weighted graph construction requires computing and storing embedding-based cosine similarities, which increases preprocessing complexity. Given the limited improvement observed here, edge weighting does not appear to be the primary factor driving GraphRAG performance in this configuration.

The retrieval strategy itself also has a weaker effect than expected. The Top-5 most similar retrieval does not clearly dominate random neighbor retrieval. In citation graphs, both retrieval strategies reach similar peak values, and both show declining behavior in several late-year horizons. In textual-similarity graphs, Top-5 retrieval is somewhat more stable in the directed setting, but the difference remains moderate. This suggests that the limiting factor may not be only the retrieval strategy, but also the ability of the smaller LLM to use the retrieved context effectively.

A notable distinction appears between citation and textual-similarity graphs. In the graph-based neural approach, citation graphs were clearly stronger than textual-similarity graphs. Here, however, the textual-similarity graphs are at least competitive and, in the later horizons, often stronger. The best late-horizon values are observed in textual-similarity settings, reaching about 0.65--0.66 AUC. This likely occurs because the LLM operates directly on textual evidence. When the graph is built from semantic similarity, the retrieved neighbors may be more topically aligned with the target paper, making the context easier for the model to interpret than citation-based neighbors alone.

Nevertheless, the overall performance remains limited. One possible explanation is that GPT-5.4 Nano is not sufficiently strong to reliably integrate the target abstract, neighbor metadata, graph configuration, and historical performance signals into a calibrated long-horizon prediction. Another possibility is that the retrieved neighbors introduce noisy or partially conflicting information, especially when selected randomly or when the graph neighborhood is not highly homogeneous. Therefore, this experiment suggests that GraphRAG performance depends not only on the graph and retrieval design, but also strongly on the reasoning and context-integration capabilities of the LLM used.

Taken together, these results show that graph-retrieved context does not automatically guarantee high predictive performance. Directionality remains important, textual-similarity graphs become relatively more competitive than in the graph-based neural setting, and weighting has limited impact. However, with a lightweight model such as GPT-5.4 Nano, the overall gains from GraphRAG retrieval are modest. This reinforces the need to evaluate retrieval-augmented pipelines under realistic model-capability and cost constraints, rather than assuming that adding retrieved context will necessarily improve prediction.

\subsubsection{Effect of graph-retrieved context on LLM prediction performance}
\label{chapter:graphragpredictionperf}

\begin{figure}[h]
\centering
\includegraphics[scale=0.6]{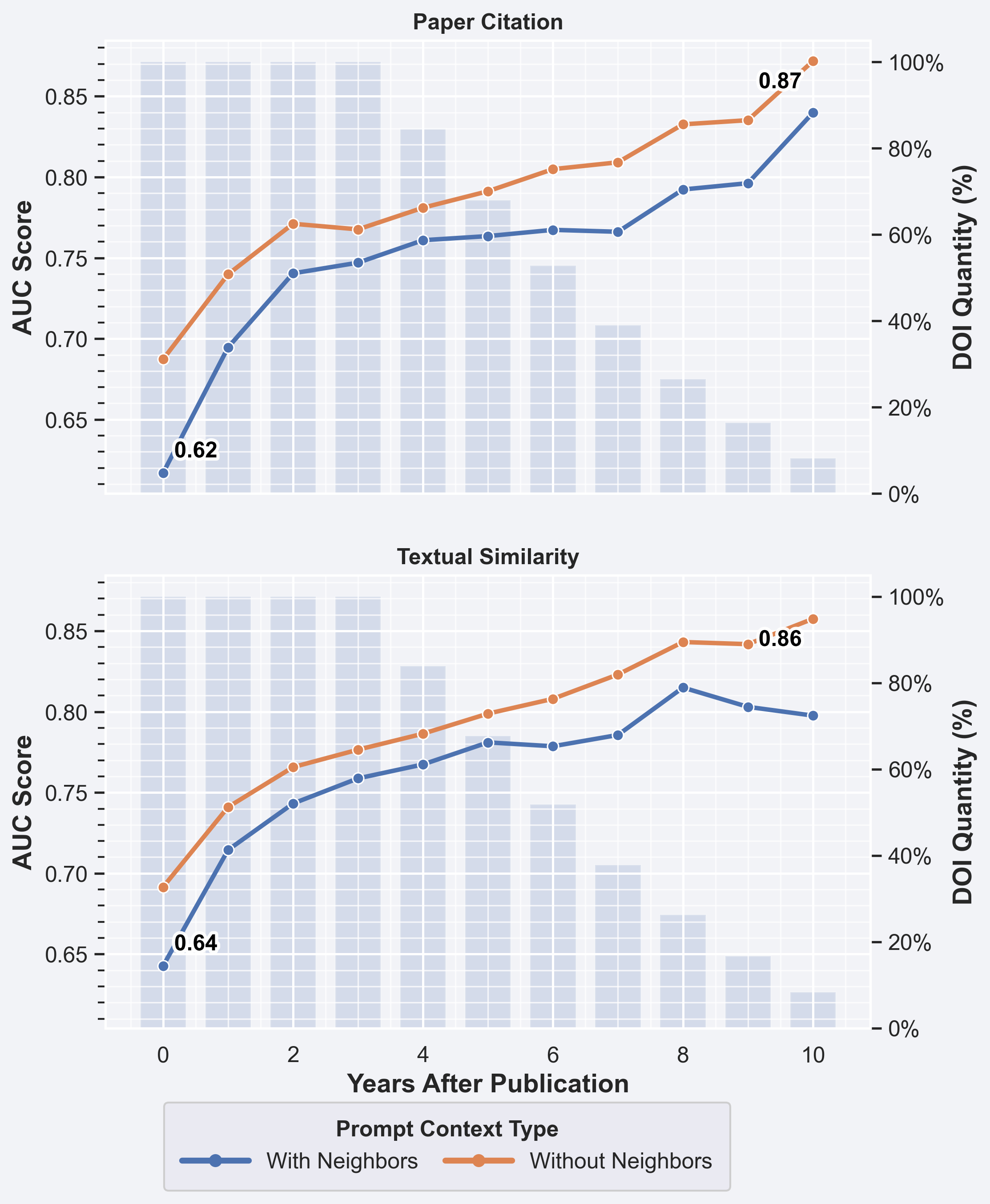}
\caption{{Effect of graph-retrieved context on LLM-based top-paper prediction.} The figure compares AUC scores over years after publication for prompts with neighbors and without neighbors, using citation and textual-similarity graphs. The background bars indicate the relative DOI quantity available at each horizon year, expressed as a percentage of the maximum number of papers in the evaluation set.}
\label{fig:LLMCompNeigh}
\end{figure}

In Figure \ref{fig:LLMCompNeigh}, we replicate the GraphRAG setup under a simplified and controlled configuration, using directed and unweighted graphs with random retrieval, and compare prompts \emph{with} and \emph{without} graph-retrieved neighbors. The goal of this experiment is to isolate the contribution of retrieved graph context itself: if the neighbors provide useful complementary evidence, the context-augmented prompt should outperform the target-only prompt. In contrast to the previous experiment with GPT-5.4 Nano, here we use GPT-5.5 with low reasoning effort, allowing us to evaluate whether a stronger model is able to make better use of the target paper and the retrieved neighborhood. The background bars show the relative number of papers available at each horizon year. As expected, the number of available papers decreases for longer horizons, since only older papers can be evaluated many years after publication. Therefore, late-horizon results should be interpreted with this reduction in effective sample size in mind.

The main result is that graph-retrieved context does not improve performance in this setting. On the contrary, the \emph{without neighbors} condition consistently outperforms the \emph{with neighbors} condition for both graph families and across all horizon years. In the citation graph, the target-only prompt starts around 0.69 AUC and rises steadily to approximately 0.87 in the final horizon, while the prompt with neighbors starts around 0.62 and reaches approximately 0.84. A similar pattern appears for the textual-similarity graph: the target-only prompt starts around 0.69 and reaches approximately 0.86, whereas the context-augmented version starts around 0.64 and remains lower throughout most of the horizon, ending around 0.80. Thus, although both configurations improve over time, the retrieved neighborhood does not provide a measurable advantage over the simpler target-only prompt.

This result is particularly relevant because the target-only LLM setting reaches performance comparable to, and in some horizons slightly above, the best graph-embedding configuration reported earlier, which achieved approximately 0.84 AUC. This suggests that GPT-5.5 can extract strong predictive signals directly from the target paper metadata, especially title, abstract, journal information, and publication year. In this case, the LLM does not appear to require additional graph-retrieved examples to produce a strong impact prediction. Rather than acting as useful evidence, the retrieved neighbors may introduce noise, ambiguity, or partially conflicting signals that make the prediction task harder.

A plausible interpretation is that the target paper already contains substantial information about its future impact potential, such as topic relevance, novelty, methodological scope, and expected audience. A stronger LLM can exploit these signals directly. The retrieved neighbors, although related through the graph, may not always be sufficiently discriminative or aligned with the target's future trajectory. This is especially important in a prediction task where the relevant distinction is not simply topical similarity, but whether a paper will become unusually influential within its cohort.

This finding has an important methodological implication for retrieval-augmented scientific prediction. GraphRAG should not be assumed to be beneficial merely because it adds more information to the prompt. In this experiment, a simpler target-only prompting strategy was both cheaper and more effective than the context-augmented version. Therefore, the value of GraphRAG depends on the quality and complementarity of the retrieved context, not simply on the presence of retrieval. In this setting, less context appears to be better: removing the neighborhood reduced potential noise and allowed the model to focus on the target paper itself.

\subsubsection{Cross-journal evaluation with and without graph-retrieved context}
\label{chapter:graphragcrossjournal}

To assess whether the behavior observed in the main corpus was specific to a single journal or reflected a more general property of the LLM-based prediction setup, we repeated the context-ablation experiment on three additional journals: \emph{Informetrics}, \emph{PNAS}, and \emph{PRL}. We kept the same controlled configuration used in the previous comparison: directed and unweighted graphs, with five randomly selected neighbors in the context-augmented condition. The purpose of this experiment was not to reproduce the entire graph-based factorial analysis across journals, but to test the external validity of the most important finding from the GraphRAG experiments: whether graph-retrieved context improves or harms LLM-based top-paper prediction.

This cross-journal evaluation was performed only for the LLM-based pipeline for methodological and practical reasons. The graph-based stage already provides a complete factorial analysis of graph construction, directionality, weighting, textual embeddings, and input representations on the main large-scale corpus. Repeating the full graph-embedding pipeline for all additional journals would require reconstructing all graph variants, recomputing {Node2Vec} embeddings, retraining the neural classifiers, and re-running the full set of horizon-based evaluations. By contrast, in the GraphRAG setting, the graph is used only as a retrieval substrate, making cross-journal evaluation substantially more direct. Therefore, the additional journals are used here specifically as an external validation of the LLM prompting and context-retrieval behavior, rather than as a full second benchmark of the graph-embedding methodology.

\begin{figure}[h]
\centering
\includegraphics[scale=0.5]{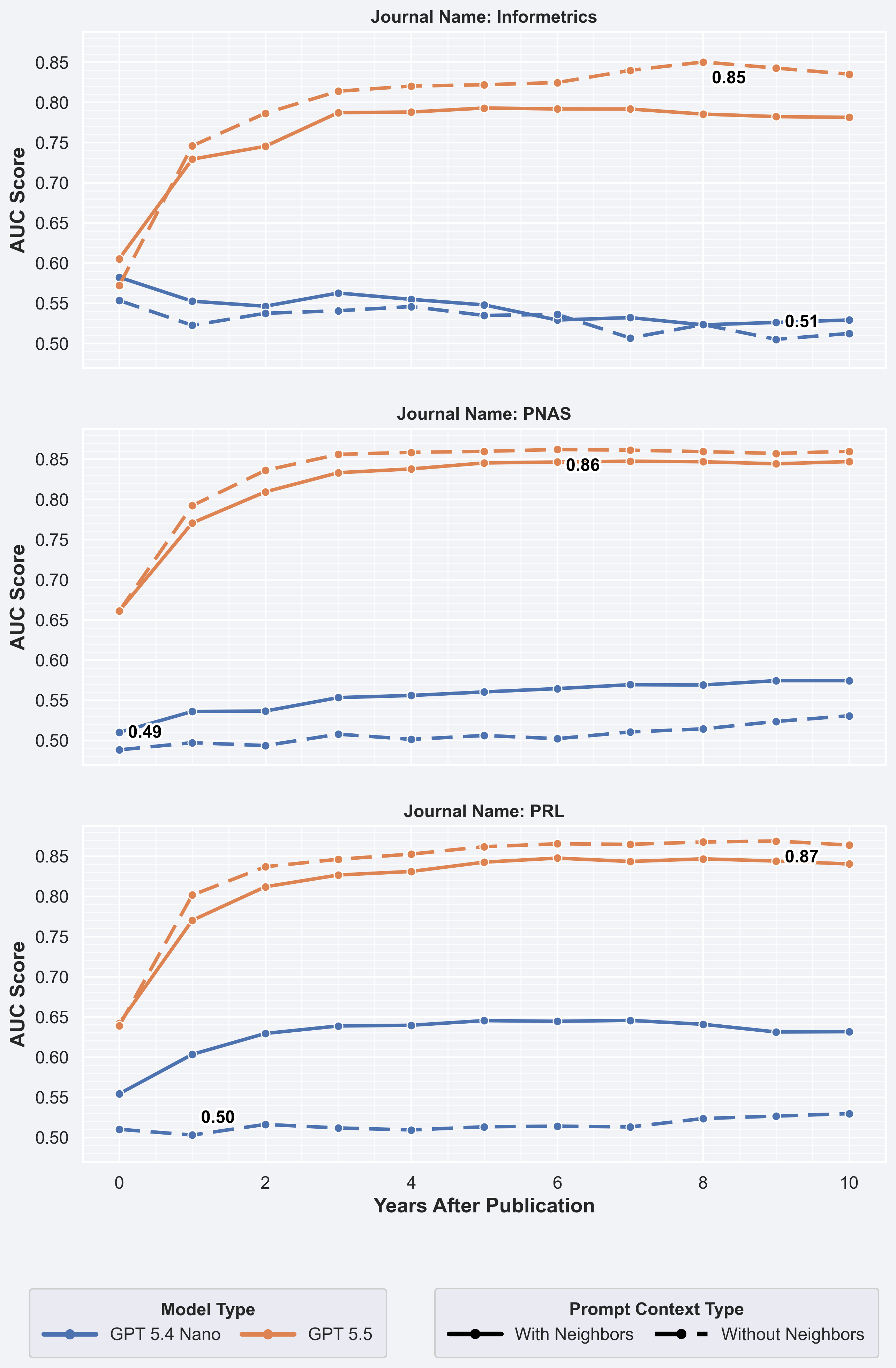}
\caption{{Cross-journal evaluation of LLM-based top-paper prediction with and without graph-retrieved context.} The figure reports AUC scores across years after publication for three journals---\emph{Informetrics}, \emph{PNAS}, and \emph{PRL}---comparing prompts augmented with graph neighbors against prompts using only the target paper. Results are shown for GPT-5.4 Nano and GPT-5.5 under the same directed, unweighted graph configuration used in the comparative GraphRAG setting.}
\label{fig:LLMExtraJournals}
\end{figure}

The results are shown in Figure \ref{fig:LLMExtraJournals}. Two patterns are immediately visible. First, model capacity matters substantially. GPT-5.5 clearly outperforms GPT-5.4 Nano across all three journals and across almost all horizon years. While GPT-5.5 rapidly reaches the 0.78--0.87 AUC range, GPT-5.4 Nano remains close to the random baseline in several conditions, with many values around 0.50--0.57 and only the strongest cases approaching approximately 0.65. This confirms that the ability to infer future scientific impact from paper metadata and retrieved context depends strongly on the reasoning and representation capabilities of the underlying LLM.

Second, the effect of graph-retrieved context depends on model capacity. For GPT-5.5, the target-only prompt, without retrieved neighbors, consistently performs better than the context-augmented prompt. This pattern appears in \emph{Journal of Informetrics}, \emph{PNAS}, and \emph{PRL}, confirming the result observed in the main corpus. In this case, adding neighbors seems to introduce noise or redundant information rather than useful complementary evidence. The strongest GPT-5.5 results reach approximately 0.85 AUC for \emph{Informetrics}, 0.86 for \emph{PNAS}, and 0.87 for \emph{PRL}, with the best values obtained without graph-retrieved context.

For GPT-5.4 Nano, however, the pattern is different. Removing the context generally weakens performance, especially in \emph{PNAS} and \emph{PRL}. This suggests that smaller models may benefit more from retrieved neighbors because they have less internal capacity or less accessible parametric knowledge to infer impact from the target paper alone. In other words, graph-retrieved context appears more useful when the LLM itself is weaker, whereas for stronger models the same context may become unnecessary or even harmful. This interaction between model capacity and retrieval usefulness is an important result: GraphRAG should not be treated as uniformly beneficial or detrimental, but as a mechanism whose value depends on the model and the task.

Another relevant observation is the fast early improvement of GPT-5.5. In all three journals, performance increases sharply within the first one or two years after publication and then stabilizes near its later-horizon values. This indicates that the target paper already contains early signals associated with future impact, but these signals become easier to identify as the first post-publication evidence accumulates. This pattern is consistent with the broader findings of this study and suggests a promising direction for future work: incorporating early dissemination indicators, such as views, downloads, online attention, or early readership metrics, to better understand how papers propagate before citation counts fully mature.

Taken together, these results suggest that the role of GraphRAG in scientific impact prediction is more nuanced than simply adding context to a prompt. For a stronger model such as GPT-5.5, the target paper alone provides enough information to achieve the best performance, and retrieved neighbors may introduce distracting or weakly discriminative evidence. For a smaller model such as GPT-5.4 Nano, however, the retrieved context can help compensate for weaker internal inference capabilities. Therefore, the usefulness of GraphRAG depends not only on the quality of the retrieved neighborhood, but also on the capacity of the LLM consuming that context. This reinforces the need to empirically validate retrieval augmentation rather than assuming that more context necessarily leads to better scientific impact prediction.

\section{Conclusion}
\label{chapter:conclusion}

In this work, we investigated scientific impact prediction through two complementary approaches: a graph-based supervised learning pipeline and an LLM-based GraphRAG-style inference framework. The common objective was to predict whether a paper would become a top paper within its journal cohort across different horizons after publication. Across both families of experiments, a consistent temporal pattern emerged: predictive performance is limited but above random at the publication year, improves sharply during the first years after publication, and then tends to stabilize or fluctuate moderately in later horizons.

In the graph-based setting, the strongest results were obtained with citation graphs, especially when structural information from {Node2Vec} was combined with OpenAI's textual embeddings. For the top-20\% classification task, the best citation-based hybrid configuration reached approximately 0.84 AUC, compared with about 0.80 when using {Node2Vec} alone. In textual-similarity graphs, the hybrid representation reached around 0.80 AUC, while the structural-only version peaked near 0.76. These findings indicate that citation structure captures important relational signals for scientific impact, while textual information acts as a strong complement, especially when it is added to a citation-based representation. We also found that edge directionality is more important than edge weighting: directed graphs, particularly citation graphs, consistently improved performance, whereas weighted edges produced only marginal and inconsistent gains.

The sensitivity analyses further clarified the behavior of the graph-based models. In textual-similarity graphs, increasing the neighborhood size from $K=3$ to $K=5$ improved performance slightly, but larger values produced little additional benefit. This suggests that $K=5$ offers a good balance between predictive performance and graph sparsity. We also found that stricter quantile thresholds make the classification task easier: the best citation-based model reached approximately 0.85 AUC under more selective top-paper definitions, while the best textual-similarity model reached around 0.82. This suggests that the proposed methodology is particularly effective for identifying highly exceptional papers rather than distinguishing moderate levels of success.

The GraphRAG-based experiments produced a more nuanced picture. When using GPT-5.4 Nano to evaluate many graph and retrieval configurations at lower cost, performance was modest, with most configurations starting around 0.53--0.56 AUC and the best textual-similarity settings reaching approximately 0.65--0.66. In this setting, directionality remained important, textual-similarity graphs became more competitive than citation graphs, and edge weighting had limited impact. However, the results also showed that graph-retrieved context alone is not sufficient when the underlying LLM has limited capacity to integrate scientific metadata, neighbor evidence, and temporal signals into calibrated predictions.

When using GPT-5.5 with low reasoning effort in the context-ablation experiment, performance improved substantially. However, the most important finding was that graph-retrieved context did not improve the stronger model's performance. In the main dataset, the target-only prompt outperformed the context-augmented prompt for both graph families, reaching approximately 0.87 AUC in the citation setting and 0.86 in the textual-similarity setting, compared with about 0.84 and 0.80 when neighbors were included. This indicates that, for a stronger LLM, the target paper's own metadata---especially title, abstract, journal information, and publication year---already contains sufficient predictive signal, while retrieved neighbors may introduce noise or weakly discriminative evidence.

The cross-journal evaluation on \emph{Informetrics}, \emph{PNAS}, and \emph{PRL} reinforced this interpretation while adding an important qualification. GPT-5.5 consistently outperformed GPT-5.4 Nano, reaching approximately 0.85, 0.86, and 0.87 AUC in the three journals, respectively. For GPT-5.5, the best results were again obtained without graph-retrieved context. For GPT-5.4 Nano, however, the opposite pattern often appeared: removing context weakened performance, suggesting that smaller models may benefit more from retrieved neighbors because they have weaker internal inference capabilities. Thus, the usefulness of GraphRAG depends not only on the quality of the retrieved context, but also on the capacity of the model consuming that context.

Overall, our results show that scientific impact prediction benefits from both structural and semantic information, but the best way to exploit these signals depends on the modeling paradigm. For supervised graph-based models, the most effective strategy is to combine citation-based structure with textual embeddings. For LLM-based prediction, graph-retrieved context is not automatically beneficial: it can help smaller models, but for stronger models the target paper alone may be more effective. These findings suggest that GraphRAG should not be treated as a default improvement, but as a design choice whose value must be validated empirically for each model, dataset, and prediction task.

Future work should explore more selective retrieval strategies, neighbor summarization, and mechanisms to filter noisy or weakly related contextual evidence before prompting the LLM. It would also be valuable to evaluate temporally controlled corpora that extend beyond the training cutoff of the LLM, in order to better isolate genuine predictive reasoning from possible parametric knowledge acquired during pretraining. Finally, incorporating early dissemination signals such as downloads, views, online attention, or readership metrics may help bridge the gap between publication-time metadata and later citation impact.

\section*{Acknowledgments}

The authors gratefully acknowledge financial support from the São Paulo Research Foundation (FAPESP) (grant no. 2025/00944-6) and the National Council for Scientific and Technological Development (CNPq-Brazil) (grant no. 304189/2025-1).

\bibliographystyle{apalike}

\bibliographystyle{abbrvnat}

\newpage

\appendix

\appendix

\section{Prompt Templates Used in the LLM-Based Experiments}
\label{appendix:prompt_templates}

This appendix presents the prompt templates used in the LLM-based prediction experiments. The prompting protocol consisted of three parts: a \emph{system prompt}, a \emph{developer prompt}, and a structured \emph{user prompt} serialized in XML format. In the final version of the experiment, the model was instructed to return only one prediction output: \texttt{y\_acc\_vector}, corresponding to the probability that the target paper becomes a top paper by accumulated citations at each requested horizon year. The prompt therefore focuses exclusively on the accumulated-citation prediction task used in the analyses reported in this work.

\subsection{System Prompt}

\begin{tcolorbox}[promptbox,title=System Prompt]
\small

\textbf{You are a scientific impact prediction engine for journal articles.}

Your job is to estimate calibrated probabilities for whether a target paper will become a \emph{top paper} within its journal at each requested horizon year.

\medskip
\textbf{Output rules}

\begin{enumerate}[leftmargin=*,itemsep=0.25em,topsep=0.4em]
    \item Output valid JSON only. No Markdown, no explanation, and no extra text.
    
    \item The JSON must have exactly one top-level key: \texttt{"response"}.
    
    \item \texttt{"response"} must have exactly one key: \texttt{"y\_acc\_vector"}.
    
    \item The required structure is:
    
    \begin{center}
    \texttt{\{"response":\{"y\_acc\_vector":[...]\}\}}
    \end{center}
    
    \item \texttt{"y\_acc\_vector"} must contain numeric probabilities in $[0,1]$.
    
    \item Probabilities must be numbers, not strings.
    
    \item The length of \texttt{"y\_acc\_vector"} must be exactly equal to \texttt{<OUTPUT\_SPEC><n\_years>}.
    
    \item Do not reveal reasoning or chain-of-thought. Return only the final JSON.
    
    \item Use only the information explicitly present in the XML input.
    
    \item Do not use external facts or hidden assumptions about papers, authors, journals, venues, identifiers, files, or datasets.
\end{enumerate}

\end{tcolorbox}

\subsection{Developer Prompt}

\begin{tcolorbox}[promptbox,title=Developer Prompt]
\small

\textbf{Task}

Predict, for the target journal article, the probability that it will be a \emph{top paper} by accumulated citations at each requested horizon year.

\medskip
\textbf{Positive event}

The positive event is defined by \texttt{<CONFIG><q\_value>}:

\begin{itemize}[leftmargin=*,itemsep=0.25em,topsep=0.3em]
    \item \texttt{q\_value} is a quantile threshold.
    \item A paper is considered \emph{top} if it belongs to the top $(1 - q\_value)$ fraction within its journal or context.
    \item Example: \texttt{q\_value = 0.8} means the positive event is being in the top 20\%.
\end{itemize}

\medskip
\textbf{Required prediction}

Return one probability for each horizon year listed in \texttt{<TARGET><years>}.

\begin{itemize}[leftmargin=*,itemsep=0.25em,topsep=0.3em]
    \item The output vector is:
    
    \begin{center}
    \texttt{y\_acc\_vector[$h$] = P(target paper is a top paper by accumulated citations up to horizon $h$)}
    \end{center}
    
    \item The order of the probabilities must follow the exact order of \texttt{<TARGET><years>}:
    \begin{itemize}[leftmargin=1.5em,itemsep=0.2em,topsep=0.2em]
        \item first probability $\rightarrow$ first year in \texttt{<TARGET><years>};
        \item second probability $\rightarrow$ second year in \texttt{<TARGET><years>};
        \item and so on.
    \end{itemize}
\end{itemize}

\medskip
\textbf{Input information}

The XML contains three main sections.

\medskip
\textbf{1) \texttt{<CONFIG>}}

This section provides the general experimental context:

\begin{itemize}[leftmargin=*,itemsep=0.25em,topsep=0.3em]
    \item \texttt{graph\_name}: name of the graph or retrieval setting;
    \item \texttt{retrieval\_type}: whether neighbors were retrieved by top-$k$ similarity/context or randomly;
    \item \texttt{K\_NEIGHBORS}: number of retrieved neighbors requested;
    \item \texttt{is\_directed} and \texttt{is\_weighted}: graph construction flags;
    \item \texttt{q\_value}: quantile threshold defining the positive class.
\end{itemize}

\medskip
\textbf{2) \texttt{<TARGET>}}

This section describes the paper to be predicted:

\begin{itemize}[leftmargin=*,itemsep=0.25em,topsep=0.3em]
    \item title;
    \item abstract;
    \item publication year;
    \item domain, field, and subfield;
    \item years: horizon years to predict;
    \item \texttt{n\_years}: exact number of probabilities required in the output;
    \item years to predict;
    \item maximum observable year.
\end{itemize}

Use the target title, abstract, publication year, field information, and requested horizons as the main basis for the prediction.

\medskip
\textbf{3) \texttt{<NEIGHBORS>}}

This section contains contextual papers retrieved from the graph:

\begin{itemize}[leftmargin=*,itemsep=0.25em,topsep=0.3em]
    \item each neighbor has metadata, text, and a \texttt{y\_acc\_vector};
    \item neighbor \texttt{y\_acc\_vector} values are historical or contextual calibration examples;
    \item they can help estimate how papers with related metadata or text behaved under the same target definition;
    \item do not copy neighbor vectors directly;
    \item do not assume the target has the same outcome as any individual neighbor;
    \item do not use neighbor vector length to decide output length.
\end{itemize}

\medskip
\textbf{Prediction guidance}

Estimate calibrated probabilities from the available evidence:

\begin{itemize}[leftmargin=*,itemsep=0.25em,topsep=0.3em]
    \item target paper metadata and text;
    \item field, domain, and subfield context;
    \item publication year and requested horizon years;
    \item neighborhood context and neighbor \texttt{y\_acc\_vector} values, when informative.
\end{itemize}

Do not force monotonicity, thresholds, labels, or fixed patterns unless supported by the input. Output probabilities, not binary decisions.

\medskip
\textbf{Leakage prevention}

\begin{itemize}[leftmargin=*,itemsep=0.25em,topsep=0.3em]
    \item The target's true \texttt{y\_acc\_vector} is not provided and must not be inferred from hidden conventions.
    \item Do not treat identifiers, graph names, retrieval order, filenames, or dataset-specific patterns as labels.
    \item If any field appears to describe the target's observed future outcome, ignore it.
\end{itemize}

\medskip
\textbf{Output validation}

Before producing the final JSON, ensure that:

\begin{itemize}[leftmargin=*,itemsep=0.25em,topsep=0.3em]
    \item there is exactly one key inside \texttt{"response"};
    \item that key is \texttt{"y\_acc\_vector"};
    \item \texttt{y\_acc\_vector} has exactly \texttt{<OUTPUT\_SPEC><n\_years>} values;
    \item every value is a numeric probability between 0 and 1;
    \item no extra text is included.
\end{itemize}

If evidence is weak or incomplete, still return a valid probability vector with exactly \texttt{<OUTPUT\_SPEC><n\_years>} values.

\end{tcolorbox}

\subsection{User Prompt Template}

The user prompt was generated dynamically from the experiment payload and serialized in XML format. It always contained a \texttt{<REQUEST>} root block. The \texttt{<OUTPUT\_SPEC>} section specified the required JSON schema and the exact output vector length. The \texttt{<CONFIG>} section described the graph and retrieval settings. The \texttt{<TARGET>} section contained the metadata and text of the target paper. In the GraphRAG condition, the \texttt{<NEIGHBORS>} section listed the retrieved contextual papers and their accumulated-citation performance vectors. In the context-free baseline, the \texttt{<NEIGHBORS>} section was omitted or left empty.

\begin{tcolorbox}[promptbox,title=User Prompt Template]
\begin{Verbatim}[fontsize=\small]
<REQUEST>

  <OUTPUT_SPEC>
    <required_json_schema>{"response":{"y_acc_vector":[...]}}</required_json_schema>
    <n_years>{n_years}</n_years>
    <required_vector_length>{n_years}</required_vector_length>
    <probability_order>same order as TARGET years</probability_order>
  </OUTPUT_SPEC>

  <CONFIG>
    <graph_name>{graph_name}</graph_name>
    <retrieval_type>{retrieval_type}</retrieval_type>
    <K_NEIGHBORS>{K_NEIGHBORS}</K_NEIGHBORS>
    <is_directed>{is_directed}</is_directed>
    <is_weighted>{is_weighted}</is_weighted>
    <q_value>{q_value}</q_value>
  </CONFIG>

  <TARGET>
    <title>{title}</title>
    <abstract>{abstract}</abstract>
    <publication_year>{publication_year}</publication_year>
    <domain>{domain}</domain>
    <field>{field}</field>
    <subfield>{subfield}</subfield>
    <years>{years}</years>
    <n_years>{n_years}</n_years>
    <years_to_predict>{years_to_predict}</years_to_predict>
    <max_year>{max_year}</max_year>
  </TARGET>

  <NEIGHBORS>
    <PAPER>
      <title>{title}</title>
      <abstract>{abstract}</abstract>
      <publication_year>{publication_year}</publication_year>
      <domain>{domain}</domain>
      <field>{field}</field>
      <subfield>{subfield}</subfield>
      <years>{years}</years>
      <y_acc_vector>{y_acc_vector}</y_acc_vector>
    </PAPER>
  </NEIGHBORS>

</REQUEST>
\end{Verbatim}
\end{tcolorbox}

\end{document}